\documentstyle[prl,aps,12pt]{revtex}
\input{epsf.sty}

\def\b\mu{{\bf \mu}}

\begin{document}

\draft

\title{\Large \bf Dressing a Scalar Mass up to Two-Loop Order\\
at Finite Temperature}
\vspace{1cm}
\author{H. C. G. Caldas\thanks{e-mail: hcaldas@funrei.br}}

\address{Departamento de Ci\^{e}ncias Naturais, DCNAT \\
Funda\c{c}\~ao de Ensino Superior de S\~{a}o Jo\~{a}o del Rei, FUNREI,\\ 
Pra\c{c}a Dom Helv\'ecio, 74, CEP:36300-000, S\~{a}o Jo\~{a}o del Rei, MG, Brazil} 

\date{October, 2001}

\maketitle                          

\vspace{0.5cm}

\begin{abstract}
In this paper we use Modified Self-Consistent Resummation (MSCR) in order to obtain the 
scalar dressed mass by evaluating the self-energy up to two loops in the neutral 
scalar $\lambda \phi^4$ model at finite temperature. With this laboratory model we show that, if a theory is renormalizable at zero temperature, by using MSCR it is always possible to obtain a finite corrected mass at finite temperature. This feature of MSCR is not observed in some other approximation techniques usually found in the literature.

\end{abstract}

\vspace{2cm}


\newpage

\section{Introduction} 

It is well known that particles immersed in a thermal medium at temperature $T$ have their properties modified. The knowledge of the reasons for these modifications and their consequences is of great relevance in the context of finite temperature field theory (FTFT). However, quantum field theory at finite temperature has some subtleties such as the breakdown of the perturbative expansion which is manifested in two cases. The first one is the appearance of infrared divergences in massless field theories or tachyonic masses in theories with spontaneous symmetry breaking \cite{Weinberg,Dolan,Linde}. The second breakdown surges in calculations using higher order diagrams where large $T$ compensates for the powers of the coupling constant. The practical way to circumvent these problems is a resummation of certain classes of graphs up to a given order in the perturbative expansion. This resummation (of an infinite set of diagrams) has been studied intensively in the last three decades \cite{Weinberg,Dolan,Cornwall,Pisarski1,Pisarski2,Kapusta,Braaten0}. Nevertheless, the use of these kind of nonperturbative methods does not altogether solve the problem of the ultraviolet (UV) divergences (which will become temperature dependent through the gap equation for the mass) that are still present. In the related literature, it is frequently claimed that the poles are suppressed by the renormalization procedure, but it is not always shown how. It is not necessary to say that, if one wants believable results, the problems raised by the divergences must be satisfactorily resolved. For example, consider the contribution to the mass squared at one loop in the neutral $\lambda \phi^4$ model, 

\begin{equation}
\label{eq0}
\Pi_1 (m_0)= \frac {\lambda}{2}~T \sum \limits_n \int \frac{d^3 k}{(2 \pi)^3} \frac{1}{\omega_n ^2 + \omega_k^2}= \frac{\lambda}{2} \int \frac{d^3 k}{(2 \pi)^3} \frac{1}{\omega_k(m_0)} 
\left[ n_B(\omega) + \frac{1}{2} \right],
\end{equation}
where $m_0$ is the vacuum mass, $\omega_{n}$ are the Matsubara frequencies, defined as $\omega_{n}=2n\pi T$ for bosons, $n_B(x)=1/(e^{\beta x}-1)$ is the usual Bose-Einstein distribution and $\omega_k(m)$, the relativistic energy, is given by $\sqrt{k^2+m^2}$. The first term in the integral is the finite temperature contribution, which is free from UV divergences. The last term in the integral is 
the zero temperature part (at this stage) and is divergent. By the use of some non-perturbative method like the Hartree approximation, one gets $M^2(T)=m_0^2+ \Pi_1 (M)$, with $M$ being the dressed mass. If the mass running in the loop is the vacuum mass (i.e., as in conventional perturbation theory), then the theory can also be renormalized at finite temperature \footnote{Considering that the theory under investigation is renormalizable at zero temperature. In the case of the $\lambda \phi^4$ model one has to consider also the ``triviality'' behavior of the theory \cite{Camelia0}.}, but if some resummation approximation is naively used as in the example above, the theory is non-renormalizable at finite temperature since one can not absorb in the {\it original} Lagrangian the infinity term, $\lambda \frac{M^2(T)}{(4 \pi)^2} \frac{1}{\tilde \epsilon}$ (obtained using dimensional regularization \cite{Pierre}). The nonobservance of this ``incompatibility'' is one of the reasons for the failure in finding appropriate renormalization conditions for a theory in study. The difficulty in obtaining renormalization prescriptions in a theory treated by a self-consistent approximation method at finite temperature is also manifested in other theories such as the linear sigma model \cite{Camelia1,Roh1,Petropoulos,Lenaghan}. Being aware of these problems, the authors of \cite{Camelia1,Camelia0,Lenaghan} proposed to treat the linear sigma model as an effective model, with the undesired result that the theory and its predictions are sensitive to the value of the finite momentum cutoff $\Lambda$.  In fact, it was known more than 20 years ago that these approximations requires temperature-dependent counterterms \cite{Baym}. Then one concludes that in order to achieve renormalization in self-consistent approximation methods, the counterterms need also to receive the benefits of resummation \cite{Mallik,Parwani,Chiku,Caldas1}. Of course it is not necessary to resum the counterterms if one employs ordinary perturbation theory at finite temperature to leading order in the perturbative expansion\cite{Ram-Mohan,Caldas} since finite temperature does not introduce any new UV divergence to the theory \cite{Kapusta}.
In a recent paper \cite{Caldas1} we presented a Modified Self-Consistent Resummation (MSCR) 
which resums higher-order terms in a nonperturbative way and cures the problem of breakdown 
of the perturbative expansion. Since this method is developed in steps, it has 
the essential features that allow the absorption of UV divergences, 
the avoidance of overcounting of diagrams and the identification of regions of validity \cite{Caldas1}. 
The main difference between the methods used in \cite{Mallik,Parwani,Chiku} and MSCR are that in \cite{Mallik} and \cite{Chiku} a mass parameter was introduced in the beginning to be determined later by some criteria and in \cite{Parwani} a known mass parameter was added (having in both cases the subtracted term treated as a new two-point interaction), whereas MSCR keep the same fundamental theory in the recalculation of the self-energy \cite{Caldas1}. It is important to point out that, although one has, in principle, the freedom of adding and subtracting mass parameters to a Lagrangian\footnote{In the classical level this procedure changes nothing. However, to finite order in perturbation theory it has consequences \cite{Weinberg}.}, one must be careful. For instance, in the case of \cite{Chiku} (where a resummation was applied in the linear sigma model at one-loop order in the perturbative expansion), this operation provided the pion and the sigma dressed masses with the same corrections. This is incorrect, since the self-energy for these fields are different, 
at least at one-loop order. It is worth to remark that even if they had added different (and correct) contributions to the sigma and pion fields, neither the $O(4)$ linear $\sigma$ model would be renormalizable nor Goldstone's theorem would be satisfied in the region of intermediate temperatures (around $T_c$, the critical temperature) \cite{Caldas1}. This is because in that temperature region, quantum fluctuation may need a more detailed description and certainly, more than one-loop in the perturbative expansion \cite{Fendley}.
In this paper we investigate the neutral scalar $\lambda \phi^4$ model at finite temperature. 
We apply MSCR, and use imaginary time formalism to calculate the self-energy up to two-loop order in the perturbative expansion. Some previous works describing two-loop calculations may be found in \cite{Parwani,Kapusta1} and more recently in \cite{Smet}. We follow the conventions of Ref. \cite{Caldas1}. The reader who is unfamiliar with FTFT may consult \cite{Kapusta} and the more recent publications \cite{Bellac,Das}. 

This paper is organized as follows. In Section II we calculate the self-energy of the $\lambda \phi^4$ model up to two-loop order and isolate the divergent and finite temperature contributions in order to proceed with the resummation and renormalization. In Section III we briefly note MSCR which will execute this resummation. We conclude in Section IV.
\section{The Laboratory Model}

Consider the following Lagrangian density describing the self-interacting 
$\lambda \phi^4$-theory:

\begin{equation}
\label{eq1}
{\cal L}={\cal L}_0(\phi)+ {\cal L}^{int}(\phi) + {\cal L}^{ct}(\phi,\lambda),
\end{equation}
where
\begin{equation}
\label{eq2}
{\cal L}_0={1\over2}\partial_\mu\phi\partial^\mu\phi-{m^2\over2}
\phi^2,
\end{equation}
the interaction Lagrangian ${\cal L}^{int}(\phi)$ is given by
\begin{equation}
\label{eq3}
{\cal L}^{int}=-{\lambda\over 4!}\phi^4,
\end{equation}
and the counterterm Lagrangian ${\cal L}^{ct}(\phi,\lambda)$ is expressed as

\begin{equation}
\label{eq4}
{\cal L}^{ct}(\phi,\lambda)= {1\over2} A(\lambda) \partial_\mu\phi\partial^\mu\phi
-{1\over2} B(\lambda) m^2 \phi^2 - {1\over 4!} C(\lambda) \phi^4. 
\end{equation}
As we shall see next, up to two-loop level the coefficient of the counterterms are 
\cite{Pierre,Parwani}

\begin{eqnarray}
\label{eq5}
A(\lambda)= - \frac{\lambda^2}{(4 \pi)^4} \frac{1}{24 \tilde \epsilon},\\
\nonumber
B(\lambda)=  \frac{\lambda}{(4 \pi)^2} \frac{1}{2 \tilde \epsilon} + 
\frac{\lambda^2}{(4 \pi)^4} \left ( \frac{1}{2 {\tilde \epsilon}^2} -  
\frac{1}{4 \tilde \epsilon}\right),\\
\nonumber
C(\lambda)=  \frac{\lambda^2}{(4 \pi)^2} \frac{3}{2 \tilde \epsilon} + 
\frac{\lambda^3}{(4 \pi)^4} \left ( \frac{9}{4 {\tilde \epsilon}^2} -  
\frac{3}{2 \tilde \epsilon}\right),
\end{eqnarray}
where we use the modified minimal subtraction $(\overline{MS})$ scheme. Through out this paper, 
we employ dimensional regularization, but we omit, for notational 
simplicity the factor $\mu^{2\epsilon}$ which multiplies $\lambda$. In the $\lambda \phi^4$ 
model, only the self-energy has to be resumed \cite{Pisarski2,Parwani}. So, the coupling constant $\lambda$ can be treated as the expansion parameter without the necessity of 
resummation. This allow us to make the calculations in terms of the renormalized $\lambda$.

Let us now compute the one and two-loop contribution to the
self-energy of $\phi$. The only contribution to the self-energy at one-loop
comes from the diagram in Fig. 1 and it is given by Eq.(\ref{eq0}).
One obtains the self-energy in the usual fashion
\cite{Kapusta},

\begin{eqnarray}
\label{eq6}
\Pi_1 (m_0)=\Pi_1^0 (m_0)+\Pi_1^{\beta} (m_0)= \\
\nonumber
\frac{\lambda ~ m_0^2}{2(4\pi)^2}  \left[- \frac{1}{\widetilde \epsilon} -1 + \ln \left(\frac{m_0 ^2}{\mu^2}\right) \right]+
\frac{\lambda}{2}\int_0^{\infty }\frac{dp ~ p^2}{2\pi^2} \frac {n_B(\omega_p)}{\omega_p(m_0)},
\end{eqnarray}
whith $\frac{1}{\widetilde \epsilon} \equiv \frac{2}{4-d}-\gamma + \ln(4\pi)$, 
where $\gamma$ is the Euler constant and $\mu$ is the renormalization scale. The divergent contribution of this graph comes only from $\Pi_1^0 (m_0)$ and will be difined as

\begin{equation}
\label{eq7}
\Pi_1^{div}(m_0) \equiv -\frac{\lambda ~ m_0^2}{2(4\pi)^2}  \frac{1}{\widetilde \epsilon},
\end{equation}
while the finite temperature contribution is given by
\begin{equation}
\label{eq7_0}
\Pi_1^{\beta} (m_0)=\frac{\lambda}{2}\int_0^{\infty }\frac{dp ~ p^2}{2\pi^2} \frac {n_B(\omega)}{\omega_p(m_0)}.
\end{equation}

The mass and vertex counterterm generated two extra diagrams to one-loop order \cite{Pierre} which are drawn in Fig. 2. These diagrams, defined as $\Pi_1^{extra^1}$ and $\Pi_1^{extra^2}$, will be necessary to cancel the infinities formed with the mixing of $T=0$ and $T \ne 0$ pieces of Fig. 3 and are easily calculated to give:

\begin{equation}
\label{eq7_1}
\Pi_1^{extra^1}=-\frac{\lambda^2 ~ m_0^2}{4(4\pi)^2}~\frac{1}{\tilde \epsilon} \left[ \frac{1}{(4\pi)^2}  \left(\frac{1} {\tilde \epsilon} -  \ln \left( \frac{m_0^2}{\mu^2} \right)  \right) + \frac{1}{2m_0} \frac{\partial}{\partial m_0} \left(I_{\beta}^{\epsilon}(m_0)\right) \right],
\end{equation}

\begin{equation}
\label{eq7_2}
\Pi_1^{extra^2}=-\frac{3 \lambda^2 ~ m_0^2}{4(4\pi)^4} \left[\frac{1}{ {\tilde \epsilon}^2}  
+\frac{1} {\tilde \epsilon} \left(1 -\ln \left( \frac{m_0^2}{\mu^2} \right)  \right) \right]
-\frac{3 \lambda^2}{4(4\pi)^2}~\frac{1} {\tilde \epsilon}~I_{\beta}^{\epsilon}(m_0),
\end{equation}
with
\begin{equation}
\label{eq7_3}
I_{\beta}^{\epsilon}(m_0) = \mu^{4 \epsilon} \int \frac{d q^{3-2\epsilon}}{(2\pi)^{3-2\epsilon} }
\frac{n_B(\omega_q)}{\omega_q(m_0)}.
\end{equation}
At two-loop order there are two contributions $\Pi_{2,1} (m_0)$ and $\Pi_{2,2} (m_0,K)$ that are depicted in Fig. 3. We note here that only the second two-loop contribution depends on $K^\mu=(k^0,\vec k)$, the external four-momentum. The ``double scoop'' diagram is

\begin{equation}
\label{eq8}
\Pi_{2,1} (m_0)=\Pi_{2,1}^0 (m_0)+\Pi_{2,1}^{\beta} (m_0)+\Pi_{2,1}^{0,\beta} (m_0),
\end{equation}
where
\begin{equation}
\label{eq8-1}
\Pi_{2,1}^0 (m_0)=\frac{\lambda^2 ~ m_0^2}{4(4\pi)^4} \left[ \frac{1}{ {\tilde \epsilon}^2} +  \frac{1}{ \tilde \epsilon} \left(1 - 2 \ln \left( \frac{m_0^2}{\mu^2} \right)  \right) \right]
 + finite~terms,
\end{equation}

\begin{equation}
\label{eq8-2}
\Pi_{2,1}^{\beta} (m_0)= - \frac{\lambda^2~I_{\beta}^{\epsilon}(m_0)}{4}  \left[ \frac{1}{2m_0} \frac{\partial}{\partial m_0} \left(I_{\beta}^{\epsilon}(m_0)\right) + \ln \left( \frac{m_0^2}{\mu^2} \right) \right],
\end{equation}

\begin{equation}
\label{eq8-3}
\Pi_{2,1}^{0,\beta} (m_0)=\frac{\lambda^2}{4(4\pi)^2} ~ \frac{1}{\tilde \epsilon}
\left[ I_{\beta}^{\epsilon}(m_0) + \frac{m_0}{2} \frac{\partial}{\partial m_0} \left(I_{\beta}^{\epsilon}(m_0) \right) \right].
\end{equation}
The ``setting sun'' diagram gives the following contributions:

\begin{equation}
\label{eq9}
Re\Pi_{2,2} (m_0,K)=\Pi_{2,2}^0 (m_0,K)+\Pi_{2,2}^{\beta} (m_0,K)+\Pi_{2,2}^{0,\beta} (m_0,K),
\end{equation}
where we have defined $\Pi_{2,2}^0 (m_0,K)$, $\Pi_{2,2}^{\beta} (m_0,K)$ and $\Pi_{2,2}^{0,\beta} (m_0,K)$ as the real parts of the functions $G_0(K_0,\vec K)$, $G_2(K_0,\vec K)$ and $G_1(K_0,\vec K)$ respectively since, accordingly with the notation of \cite{Parwani}, $\Pi_{2,2} (m_0,K)=G_0+G_1+G_2$ . These real retarded parts have been obtained after the necessary analytic continuation $k_0 \to -i\Omega + \epsilon$, with $\epsilon \to 0^+$ and $\Omega > 0$. Now, with $K^2=\Omega^2+\vec k ^2$, one has

\begin{equation}
\label{eq9-1}
\Pi_{2,2}^0 (m_0,K^2)= \frac{\lambda^2 ~ m_0^2}{4(4\pi)^4} \left[ \frac{1}{ {\tilde \epsilon}^2} +  \frac{1}{ \tilde \epsilon} \left(3 - 2\gamma -2 \ln \left( \frac{m_0^2}{\mu^2} \right)  \right) \right]+\frac{\lambda^2}{(4 \pi)^4} \frac{K^2}{24 \tilde \epsilon} + finite~terms,
\end{equation}

\begin{equation}
\label{eq9-2}
\Pi_{2,2}^{\beta} (m_0,i \Omega,\vec k =0)=-\frac{\lambda^2} {8(2\pi)^4}  \int_0 ^{\infty}dp ~p~\frac{n_B(\omega_p)}{\omega_p(m_0)} 
 \int_0 ^{\infty} dq ~q~ \frac{n_B(q)}{\omega_q(m_0)}\ln \left|\frac{Y_+}{Y_-} \right|,
\end{equation}
where
\begin{eqnarray}
\label{eq9-21}
Y_{\pm}=[\Omega^2-(\omega_p+\omega_q+\omega_{p \pm q})^2]
[\Omega^2-(\omega_q-\omega_p+\omega_{p \pm q})^2] \\
\nonumber
\times [\Omega^2-(\omega_p-\omega_q+\omega_{p \pm q})^2]
[\Omega^2-(\omega_q+\omega_p-\omega_{p \pm q})^2],
\end{eqnarray}

\begin{equation}
\label{eq9-3}
\Pi_{2,2}^{0,\beta} (m_0,i \Omega, \vec k=0)= -\frac{\lambda^2 ~  I_{\beta}^{\epsilon}(m_0)} {2(4\pi)^2}  
\left[\frac{1}{\tilde \epsilon} - \ln \left( \frac{m_0^2}{\mu^2} \right) +2 \right] + 
F(\Omega^2),
\end{equation}
where $F(\Omega^2)$ is 

\begin{equation}
\label{eq9-4}
F(\Omega^2)= -\frac{\lambda^2} {8(2\pi)^4}  \int_0 ^{\infty}dk ~k~\frac{n_B(\omega_k)}{\omega_k(m_0)} 
 \int_0 ^{\infty}\frac{dq}{\omega_q(m_0)}\left[q \ln \left|\frac{X_+}{X_-} \right|-4k \right],
\end{equation}
with
\begin{equation}
\label{eq9-5}
X_{\pm}(\Omega^2)=[\Omega^2-(\omega_p+\omega_q+\omega_{p \pm q})^2]
[\Omega^2-(\omega_q-\omega_p+\omega_{p \pm q})^2].
\end{equation}

In Eqs. (\ref{eq8-3}) and (\ref{eq9-3}), $\Pi_{2,1}^{0,\beta} (m_0)$ and $\Pi_{2,2}^{0,\beta} (m_0)$ contain the mixed temperature-dependent divergent contributions from the $T=0$ part from one loop with the $T \ne 0$ from the second loop in each graph of Fig. 3. When the extra diagrams are taken into account, these spurious (not absorbable by counterterms) temperature-dependent divergences fortunately cancel as the reader can easily check combining Eqs. (\ref{eq8-3}) and (\ref{eq9-3}) with equation (\ref{eq7_2}). In addition, the reader can see that 
the term $\ln \left( \frac{m_0^2}{\mu^2}\right)$ in the simple pole has been also canceled \cite{Pierre}.

Thus, the divergent contribution to the mass at two-loop order after the cancelation of the divergences from the extra diagrams with the ones from $\Pi_{2,1} (m_0)$ and $\Pi_{2,2} (m_0,K)$, is given by

\begin{equation}
\label{eq10}
\Pi_2^{div}(m_0) = - \frac{\lambda^2~m_0^2}{(4 \pi)^4}  \left ( \frac{1}{2 {\tilde \epsilon}^2} -  \frac{1}{4 \tilde \epsilon}\right).
\end{equation}

The term $\frac{\lambda^2}{(4 \pi)^4} \frac{K^2}{24 \tilde \epsilon}$ in Eq.(\ref{eq9-1}) is canceled by the usual two-loop wave function renormalization counterterm whose coefficient is given by $A(\lambda)$ in Eq.(\ref{eq5}). The complete divergent contribution up to two-loop order to the squared mass is then expressed as

\begin{eqnarray}
\label{eq11}
\Pi^{div} (m_0) \equiv \Pi_1^{div}(m_0)+\Pi_2^{div} (m_0)= \\
\nonumber
-\frac{\lambda}{2(4\pi)^2}  \frac{m_0^2}{\widetilde \epsilon}
- \frac{\lambda^2}{(4 \pi)^4} m_0^2 \left ( \frac{1}{2 {\tilde \epsilon}^2} -  \frac{1}{4 \tilde \epsilon}\right),
\end{eqnarray} 
which justifies $B(\lambda)$ in Eq.(\ref{eq5}). An important remark is now in order. As we shall see in the next section, with MSCR the counterterms that remove divergences in the self-energy at zero temperature also remove the divergences at finite temperature. The only difference is that the recalculation of the self-energy changes the vacuum mass by a thermal mass. However, the cancellation of the temperature-dependent infinities is still guaranteed due to the consistence of the MSCR method. 

The temperature-dependent part of the self-energy at one and two-loop order at zero three-momentum is given by

\begin{eqnarray}
\label{eq12}
\Pi^{\beta} (m_0,i \Omega,\vec k=0) \equiv \Pi_1^{\beta}(m_0)+ \Pi_{2,1}^{\beta} (m_0) +
\Pi_{2,2}^{\beta} (m_0,i \Omega,\vec k =0) - \\
\nonumber
\frac{\lambda^2~I_{\beta}^{\epsilon}(m_0) } {2(4\pi)^2} \left[2-\ln \left( \frac{m_0^2}{\mu^2} \right) \right] + F(\Omega^2),
\end{eqnarray}
where the two terms in the second line of Eq.(\ref{eq12}) come from $\Pi_{2,2}^{0,\beta} (m_0,i \Omega, \vec k=0)$.

\section{Dressing the Mass}
\label{Dress}

We now begin our calculation of the dressed mass up to two-loop order in the perturbative expansion. The thermal mass is defined to be the real part of the pole of the corrected propagator at zero ($\vec k$) momentum

\begin{eqnarray}
\label{eq13}
{\cal D}^{-1} = {\cal D}_0^{-1} + \Pi (m_0,i \Omega,\vec k=0)=0 ~\to~\\
\nonumber
\Omega^2=m_0^2+\Pi (m_0,i \Omega,\vec k=0),
\end{eqnarray}
where ${\cal D}_0$ is the tree-level propagator and $\Pi=\Pi^0 + \Pi^{\beta}$ is the self-energy at finite temperature up to a given number of loops at zero momentum. In this work $\Pi^0$ is given by Eq.(\ref{eq11}) and $\Pi^{\beta}$ is given by Eq.(\ref{eq12}). Namely, the finite terms of $\Pi_1^0$, $\Pi_{2,1}^0$ and $\Pi_{2,2}^0$ has been dropped.

However, as we have discussed earlier, the perturbative expansion breaks down in the temperature dependent part of the self-energy mainly in two ways. The necessity of a consistent resummation is evident both at low and high temperatures. Of course this breakdown is also manifested in the $\lambda \phi^4$ model as well. The first necessity surges when there is a symmetry breaking, and in this situation we would have for the shifted-$\lambda \phi^4$ model: $m_0^2 \to m^2 = m_0^2 + \frac{1}{2} \lambda \nu^2(T)$, where $\nu(T)$ is the thermal expectation value of $\phi$ defined in $T=0$ as $\sqrt{-\frac{6m_0^2}{\lambda}}$, with $m_0^2<0$ to allow symmetry breaking. In this case, the mass running in the loops become tachyonic even below $T_c$ since $\nu(T)$ decreases as $T$ increases. The second necessity is due to the fact that higher order diagrams are larger than the lower ones at high T, even if the strength of the coupling is small. For instance, consider the diagram $ \Pi_{2,1}^{\beta} (m_0)$. In the high temperature limit, it gives $\lambda^2\frac{T^3}{m_0}$. For a diagram with $j$-bubbles attached, the power of $\frac{T}{m_0}$ become more severe, as $\lambda^j \frac{T^{2j-1}}{m_0^{2j-3}}$. Thus, higher order terms must be resummed to get sensible results at high T.

With MSCR, the resummation is consistently achieved by the recalculation of the self-energy. 
Let us now review our procedure which resums higher loop diagrams in the tree-level propagators. The goal is to make renormalization possible when obtaining the pole of the effective propagator. The method consist in recalculating the self-energy, in steps, using in each step the mass obtained in the previous one such that $ M_n ^2= (A_n+1)M_{n-1} ^2 + \Pi(M_{n-1}) $, where $n$ 
is the order of the nonperturbative correction and $A_n$ is the coefficient of the appropriate counterterm. With this procedure it is easier to identify and absorb the divergent parts of the self-energy (since the masses which multiply the divergences are necessarily the same as in counterterms) in order to have finite thermal masses. 

{\bf Step 1:}

Start with the effective Lagrangian where the mass is given by:

\begin{equation}
\label{eq14}
M_0^2=m_0^2
\end{equation}

{\bf Step 2:}

Evaluate the one and two-loop self-energy corrections to this mass from 
Eqs.(\ref{eq11}) and (\ref{eq12}) and define the first order corrected mass as

\begin{eqnarray}
\label{eq15}
M_1^2=M_0^2+ \Pi (m_0, \Omega,\vec k=0)=(A_1+1)m_0^2+\Pi (m_0, M_0,\vec k=0)= \\ \nonumber
m_0^2+\Pi^{Ren} (m_0, M_0,\vec k=0),
\end{eqnarray}
where $A_1=\frac{\lambda}{(4 \pi)^2} \frac{1}{2 \tilde \epsilon} + 
\frac{\lambda^2}{(4 \pi)^4} \left ( \frac{1}{2 {\tilde \epsilon}^2} -  
\frac{1}{4 \tilde \epsilon}\right)$ is given by the second line of Eq.(\ref{eq5}).

{\bf Step 3:}

Now we take the mass computed in the previous step and improve the results by defining a next-order nonperturbative correction. With this we get a new effective Lagrangian where the mass is given by

\begin{eqnarray}
\label{eq16}
M_2^2=M_1^2+ \Pi (M_1, \Omega,\vec k=0)= \overbrace {m_0^2+\Pi^{Ren} (m_0, M_0,\vec k=0)}^{M_1^2} +\Pi (M_1, \Omega=M_1,\vec k=0)= \\
\nonumber
(A_2+1)m_0^2+(B_2+1)\Pi^{Ren} (m_0, M_0,\vec k=0)+\Pi (M_1, \Omega=M_1,\vec k=0)= \\
\nonumber 
m_0^2+\Pi^{Ren} (M_1, \Omega=M_1,\vec k=0),
\end{eqnarray}
where $A_2=A_1$. The coefficient of the temperature dependent mass counterterm $B_2$ is fixed 
in a manner to cancel not only the divergence proportional to $\Pi (m_0, \Omega,\vec k=0)$, 
but also this term together. Generalizing, at each stage of the procedure, for $n > 1$, in 
the expressions for $M_{n}$, the self-energy $\Pi(M_{n-2})$ have to be cancelled to avoid the overcounting of diagrams. This implies that $B_2=A_2-1$. It is shown that this first recalculation (iteration) corresponds to a daisy sum \cite{Dolan,Kapusta,Caldas1}.

{\bf Step 4:} 

Proceeding with the iteration, in the limit $n \to \infty$ the mass $M_n$ 
have formally the same expression as the mass $M_{n-1}$ which is already renormalized. 
Thus, in this limit we will have, 

\begin{equation}
\label{eq17}
M_n^2 = m_0^2+ \Pi^{Ren} (M_n,\vec k=0).
\end{equation}
At each intermediate step, in the loops we set $\Omega^2=M_{n-1}^2$ in the computation of $M_{n}^2$. This ensures the cancellation of the divergences in all stages of the process, 
since the masses in the counterterms will necessarily be the same as in the divergences. 
In the end, in the resulting integral equation of interest, $\Omega^2=M^2$ as it should. 
A part of the complete self-energy diagrams in the superdaisy sum is shown in Fig. 4. 
As we are concerned only about finite temperature effects, 
$\Pi^{Ren} (M_n,\vec k=0)$ is given solely by Eq. \ref{eq12}. Strictly 
speaking, $M^2$ in Eq. (\ref{eq17}) should be obtained numerically since 
$\Pi^{Ren} (M)$ cannot be evaluated in a closed form. However, we can 
get a rough expression for the gap equation in the high temperature limit

\begin{eqnarray}
\label{eq17_1}
M^2 = m_0^2 + \frac{\lambda T^2}{24} \left(1- \frac{3M}{\pi T} \right) 
+ \frac{\lambda^2 T}{32 \pi M} \left(\frac{T^2}{12}- \frac{MT}{4 \pi} \right) \\
\nonumber
- \frac{\lambda^2 T^2}{12(4 \pi)^2} \left(1- \frac{3M}{\pi T} \right) + 
\frac{\lambda^2 T^2}{24(4 \pi)^2} \left[ \ln \left( \frac{M}{T}\right)^2 
+ 5.0669... \right], 
\end{eqnarray}
where the last term in the equation above is the computation of 
$\Pi_{2,2}^{\beta}(M) + F(M)$ which we borrow from \cite{Parwani}. 
Thus, the resummed thermal mass up to two-loop is the positive root 
of the equation

\begin{equation}
M^3 - \left[m_0^2 + \frac{\lambda T^2}{24} +
\frac{\lambda^2 T^2}{24(4 \pi)^2} \ln \left( \frac{M}{T}\right)^2 \right]M 
+ \frac{\lambda T}{8 \pi}\left(1-\frac{2 \lambda}{(4 \pi)^2} \right)M^2 
- \frac{\lambda^2 T^3}{384 \pi}=0.
\label{eq17_2}
\end{equation}

We have shown that \cite{Caldas1} to one loop-order the thermal mass of the massless $\lambda \phi^4$ in the weak coupling limit ($\lambda << 1$) at high temperature can be dressed algebraically. This simple application exemplify that, indeed, the method works very well. The main results are quoted here:

\begin{eqnarray}
\label{eq18}
M_0=0, \\
\nonumber
M_1 ^2 = M_0 ^2 + \Pi(M_0)= \frac{\lambda T^2}{24},
\end{eqnarray}

\begin{eqnarray}
\label{eq19}
M_2 ^2 = M_1 ^2 + \Pi(M_1)= (A_2 + 1)\Pi^{Ren}(M_0)+ \Pi(M_1)=\\
\nonumber
\frac{\lambda T^2}{24} \left(1- \frac{3M_1}{\pi T} \right)=\\
\nonumber
\frac{\lambda T^2}{24} \left[1- 3\left(\frac{\lambda}{24\pi^2}\right)^{\frac{1}{2}} \right],
\end{eqnarray}
The mass dressed by the second iteration, $M_2$, which was obtained in our method evaluating Fig. 1 with $M_1$ in that loop can equivalently be achieved summing an infinite set of ``daisy'' diagrams with $M_0$ in the loops. In this case all ``daisy'' diagrams are IR-divergent since $M_0=0$, but their sum is IR finite\cite{Kapusta,Parwani,Caldas1}. 
Continuing the iterations,

\begin{eqnarray}
\label{eq20}
M_3 ^2 = M_2 ^2 + \Pi(M_2)=\\
\nonumber
\frac{\lambda T^2}{24} \left[1- \frac{3M_1}{\pi T}\left(1-\frac{3M_1}{\pi T} \right)^{\frac{1}{2}} \right]=\\
\nonumber
\frac{\lambda T^2}{24} \left[1- 3 \left(\frac{\lambda}{24\pi^2}\right)^{\frac{1}{2}} + \frac{9}{2}\left(\frac{\lambda}{24\pi^2}\right)\right],
\end{eqnarray}
and finally, for the $n$-th iteration, we obtain $M_n ^2 = \frac{\lambda T^2}{24} \left\{1+ \sum_{j=1}^{n} \frac{1}{2^{j-1}} \left[-3\left(\frac{\lambda}{24\pi^2}\right)^{\frac{1}{2}} \right]^j \right\}$. The ``superdaisy'' sum (which corresponds to the limit $n \to \infty$) 
gives

\begin{equation}
\label{eq21}
M ^2 = \frac{\lambda T^2}{24} \left[ \frac{1-\frac{3}{2}\left(\frac{\lambda}{24\pi^2}\right)^{\frac{1}{2}}} {1+\frac{3}{2} \left(\frac{\lambda}{24\pi^2}\right)^{\frac{1}{2}}} \right].
\end{equation}
As one can see, the nonperturbative nature of the procedure leaves its signature in the nonanalyticity of the coupling constant in all stages of the process for $n>1$.

\section{Concluding Remarks}
\label{conc} 

In this paper we have computed the self-energy of the neutral $\lambda \phi^4$ model up to two-loop order at finite temperature. We have used MSCR in order to overcome the problems raised at finite temperature which demand resummation and renormalization. Thus, the finite thermal mass has been consistently dressed up to two-loop order in the perturbative expansion and infinitely many other diagrams nonperturbatively. A comparison with other results in the recent literature may be done. In the second reference of \cite{Kapusta1} the self-energy of a model having cubic and quartic interactions has been calculated at two-loop order. The model was used to illustrate the connection between multiloop self-energy diagrams and multiple scattering in a medium. The renormalization was discussed only for the first calculation of the self-energy i.e., with the vacuum mass $m$ in the loops though the replacement of $m$ by a thermal mass $m_T$ in the loops is considered. In \cite{Smet} the 2PPI expansion has been used to compute the effective potential and the effective mass at two-loop in the $\lambda \phi^4$ model. Although they obtain renormalized quantities, there are mainly two differences. The first is that the corrections the 2PPI expansion take into account are not the complete two-loop corrections. The double scoop diagram is missing. As discussed in Ref.\cite{Kapusta}, in order to respect the symmetries of the Lagrangian, one must retain all diagrams to the given number of loops. The second difference is that they calculated the setting sun diagram at zero four-momentum. By definition this does not give the pole and necessarily the physical mass. Our work is more closely related with Parwani's resummed perturbative expansion \cite{Parwani}. He was concerned in obtaining consistently 
the pole of the effective propagator up to order $g^4~(\lambda^2)$ taking 
into account temperature dependent counterterms as we did. The basic 
difference between his work and this one is that he used a mass corrected 
only by the one-loop diagram (which corresponds to our $M_2^2$, Eq. 
(\ref{eq19})) in the calculation of the thermal mass up to two-loop. 
However, this assured the computation of the thermal mass up to order 
$g^3$, as he intended.

Although this model is very simple, the formalism applied here ought to be used in more realistic theories since the interaction and characteristics present here may find a place there. 
A next step could be the application of some properties of the MSCR in (more involved) gauge theories.

\section*{Acknowledgements}
This work is dedicated to my son, Heron Jr., who was born in July of 2001.


\newpage

Figure Captions

Figure 1: The one-loop contribution to the self-energy.

Figure 2: The ``extra'' one-loop diagrams: (a) the mass counterterm diagram and (b) the vertex counterterm diagram.

Figure 3: Two-loop contribution to the self-energy: (c) the ``double scoop'' diagram and (d) the ``setting sun'' diagram.

Figure 4: A part of the complete self-energy diagrams in the superdaisy sum. 
The extra diagrams have been omitted.

\end{document}